\documentclass{article}
\usepackage{hiph-art}
\volnumber{19} \issuenumber{1} \edyear{2005}                             
\frompage{000} \topage{000}                                              
\recrevdate{12 October 2005}                                              
\def\rightmark{Local thermalization}
\def\leftmark{G. Wolschin, M.
Biyajima, T. Mizoguchi, and N. Suzuki} 
\title{Local Thermalization in ${d}$ + Au collisions} 
\authors{ 
{G. Wolschin$^{1,2}$, M.
Biyajima$^{2}$, T. Mizoguchi$^{3}$, and N. Suzuki$^{4}$  %
\index{Wolschin, G., Biyajima, M., Mizoguchi, T., Suzuki, N.} 
}\\[2.812mm]
{\normalsize
\hspace*{-4pt} $^{1}$ Theoretical Physics, Heidelberg University,
D-69120 Heidelberg, Germany\\
$^{2}$ Department of Physics, Shinshu University, Matsumoto 390-8621,
Japan\\
$^{3}$ Toba National College of Maritime Technology, Toba 517-8501, Japan\\
$^{4}$ Department of Comprehensive Management, Matsumoto University,\\
Matsumoto 390-1295,
Japan\\[0.2ex] 
}} 
\abstract{The extent of a locally equilibrated parton plasma in ${d}$ + Au 
collisions
at $\sqrt{s_{NN}}$ = 200 GeV is investigated as a 
function of centrality in a nonequilibrium-statistical 
framework. Based on a three-sources model, analytical solutions
of a relativistic diffusion equation are in precise agreement with recent 
data for charged-particle pseudorapidity distributions. 
The moving midrapidity source indicates the size of the local
thermal equilibrium region after hadronization. In central ${d}$ + Au 
collisions it contains 19\% of the produced particles.}
\makeindex
\begin{document}
\maketitle
\section{Introduction}\label{intro}
In order to investigate analytically the gradual thermalization occuring in the
course of particle production at the highest available energies
in heavy-ion collisions, we propose nonequilibrium-statistical 
methods. The
approach is tailored to identify the fraction of produced particles in local 
thermal equilibrium from their distribution functions in pseudorapidity.
It yields indirect evidence for the extent and
system-size dependence of a locally equilibrated parton plasma. 

Recently pseudorapidity
distributions of primary charged particles have become available
\cite{bbb05} as functions of centrality in ${d}$ + Au collisions at
a nucleon-nucleon center-of-mass energy of 200 GeV. They are investigated
within a nonequilibrium-statistical framework that is based on
analytical solutions of a Relativistic Diffusion Model (RDM), and the
results for this very asymmetric system are compared to 
Au + Au at the same nucleon-nucleon center-of-mass energy, where 
the formation of a locally equilibrated subsystem appears
to be more likely. 
\section{Relativistic Diffusion Model}\label{rdm}
Our analytical investigation is based on a linear  
Fokker-Planck equation (FPE) 
for three components $R_{k}(y,t)$ of the distribution function
in rapidity space
\cite{wol99,biy02,wol03,biy04,wbs05} 

\begin{equation}
\frac{\partial}{\partial t}R_{k}(y,t)=
\frac{1}{\tau_{y}}\frac{\partial}
{\partial y}\Bigl[(y-y_{eq})\cdot R_{k}(y,t)\Bigr]
+\frac{\partial^2}{\partial^{2} y}\Bigl[D_{y}^{k}
\cdot R_{k}(y,t)\Bigr]
\label{fpe}
\end{equation}\\
with the rapidity $y=0.5\cdot ln((E+p)/(E-p))$.
The diagonal components $D_{y}^{k}$ of the diffusion tensor  
contain the microscopic
physics in the respective Au-like (k=1), ${d}$-like (k=2)
and central (k=3) regions. They 
account for the broadening of the distribution 
functions through interactions and particle creations. 
In the present investigation the off-diagonal terms of the
diffusion tensor are assumed to be zero.
The rapidity relaxation time $\tau_{y}$ determines
the speed of the statistical equilibration in y-space.

As time goes to infinity, the mean values of the
solutions of Eqs. (\ref{fpe}) approach the equilibrium value $y_{eq}$. 
We determine it 
from energy- and momentum conservation \cite{bha53,nag84}
in the system of Au- and ${d}$-participants and hence, it 
depends on impact parameter. This dependence is decisive 
for a detailed description of the measured charged-particle
distributions in asymmetric systems:

\begin{equation}
y_{eq}(b)=1/2\cdot ln\frac{<m_{1}^{T}(b)>exp(y_{max})+<m_{2}^{T}(b)>
exp(-y_{max})}
{<m_{2}^{T}(b)>exp(y_{max})+<m_{1}^{T}(b)>exp(-y_{max})}
\label{yeq}
\end{equation}\\
with the beam rapidities y$_{b} = \pm y_{max}$ and the mean transverse
masses $<m_{1,2}^{T}(b)>$ 
that depend on the impact parameter $b$. The average 
numbers of participants $N_{1,2}(b)$
in the incident gold and deuteron nuclei are calculated from the
geometrical overlap. The results are consistent with the Glauber
calculations reported in \cite{bbb05} which we use in the further
analysis. The corresponding equilibrium values of the rapidity
vary from y$_{eq}=$ - 0.169 for peripheral (80-100$\%$) to 
y$_{eq}=$ - 0.944 for central (0-20$\%$) collisions.
They are negative due to the net longitudinal momentum of the
participants in the laboratory frame, and their absolute
magnitudes decrease with impact parameter since the number of
participants decreases for more peripheral collisions.

The FPE can be solved analytically in the linear case  
with constant $D_{y}^{k}$.
The initial conditions for produced hadrons are taken as
$\delta$-functions at the beam
rapidities, supplemented by a source centered at the equilibrium value
y$_{eq}$. This value is equal to zero
for symmetric systems, but for the asymmetric ${d}$ + Au case its
deviation from zero according to Eq.(\ref{yeq}) is decisive 
in the description of particle production.

With $\delta-$function initial conditions for the Au-like source (1),
the ${d}$-like source (2), and the equilibrium source (eq), we obtain 
exact analytical diffusion-model solutions as incoherent
superpositions of the distribution functions $R_{k}(y,t)$ because the
differential equation is linear. The total number of charged particles 
in each centrality bin
$N_{ch}^{tot}$ is determined from the data. The average number
of charged particles in the equilibrium source $N_{ch}^{eq}$ is a
free parameter that is optimized together with the variances
and $\tau_{int}/\tau_{y}$ in a $\chi^{2}$-fit of the data
using the CERN minuit-code.
\section{Comparison with RHIC-data}\label{dat}
For central collisions (0-20\%) of ${d}$ + Au, the charged-particle yield is
dominated by hadrons produced from the Au-like source, but there
is a sizeable equilibrium source that is more important
than the ${d}$-like contribution. This thermalized source is moving since
y$_{eq}$ has a finite negative value for ${d}$ + Au, whereas it is at
rest for symmetric systems.
The total yield is
compared to PHOBOS data \cite{bbb05} which refer to the
pseudorapidity $\eta=-ln[tan(\theta / 2)]$ since particle 
identification was not available.
As a consequence, there is a small difference to the model result
in $y$-space ($y\approx \eta$) which is most pronounced in the 
midrapidity region. It is removed when
the theoretical result is converted to $\eta$-space  
through the Jacobian
\begin{equation}
J(\eta,\langle m\rangle/\langle p_{T}\rangle) 
 = \cosh({\eta})\cdot [1+(\langle m\rangle/\langle p_{T}\rangle)^{2}
+\sinh^{2}(\eta)]^{-1/2}.
\label{jac}
\end{equation}
Here we approximate the average mass $<m>$ of produced charged hadrons in the
central region by the pion mass $m_{\pi}$, and use a
mean transverse momentum $<p_{T}>$ = 0.4 GeV/c.
\begin{figure}[htb]
\vspace*{-.4cm}
                 \insertplot{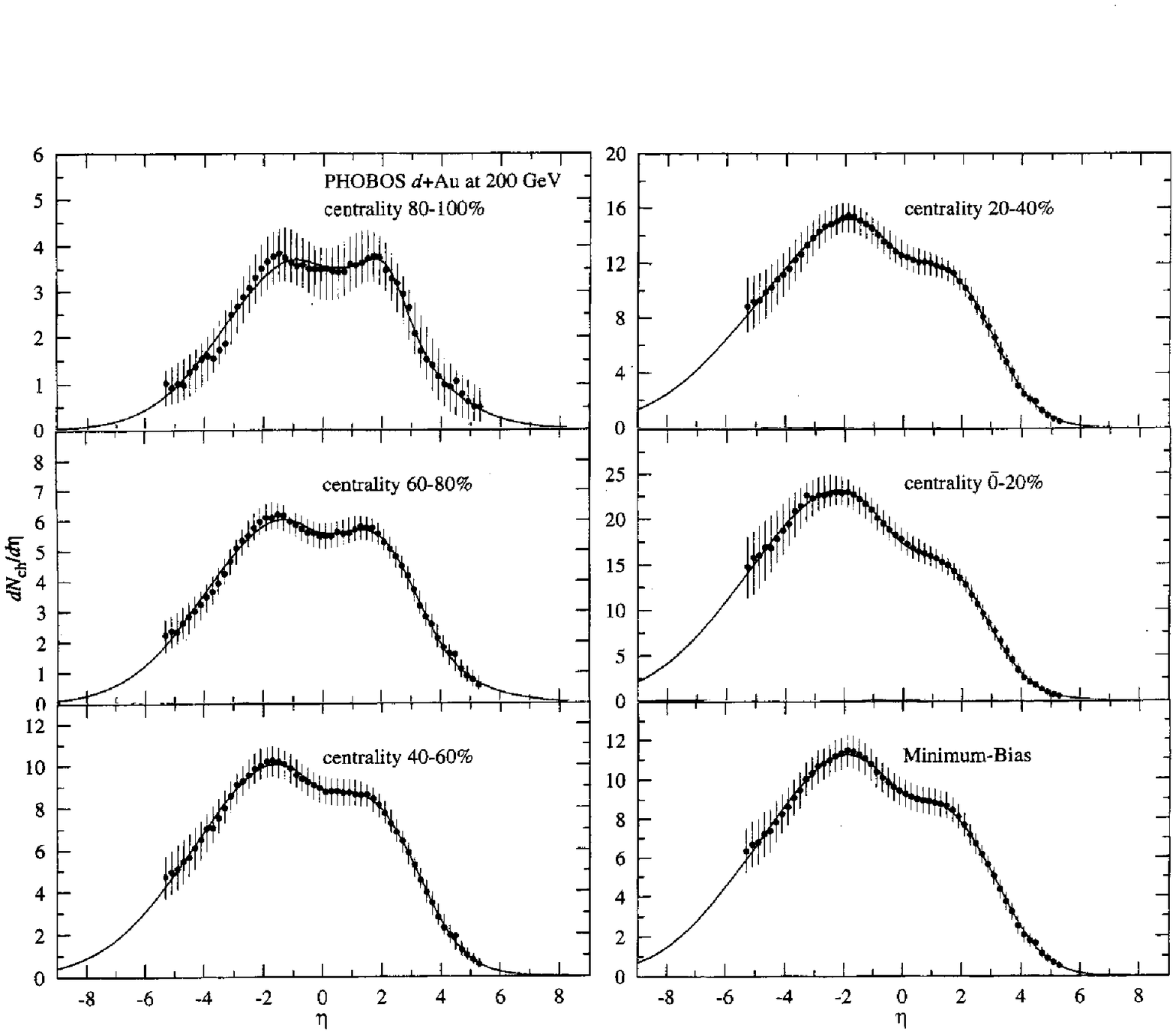}
\vspace*{-2.0cm}
\caption[]{Charged-hadron pseudorapidity spectra in the 3-sources 
Relativistic Diffusion 
Model (RDM), solid curves \cite{wbs05}, compared to $d$ + Au PHOBOS data.}

\label{fig1}
\end{figure}

The model calculations are converted to $\eta$-space and 
compared with PHOBOS data for 
five centrality cuts \cite{bbb05} and minimum bias \cite{bbb04}
in Figure 1. The minimization procedure yields precise results
so that reliable values for the relative importance of the
three sources for particle production, and for
$\tau_{int}/\tau_{y} (\simeq 0.4$ in central collisions)
can be determined \cite{wbs05}.
The observed shift of the distributions towards
the Au-like region in more central collisions, and the steeper slope 
in the deuteron direction as compared to the gold direction
appear in the Relativistic Diffusion Model as a
consequence of the gradual approach to equilibrium.

The magnitude of the equilibrium source in central ${d}$ + Au
collisions at the highest 
RHIC energy is about 19\% of the total yield. Comparing this
with a previous result \cite{biy04} for Au + Au in the 
three-sources-RDM \cite{wol03,biy04}, we note that the  
equilibrium source for 
particle production tends to be larger in the heavy
system. However, it turns out that
the determination of the number of particles in the midrapidity 
source is not unique for symmetric systems.

\section{Conclusion}\label{con}
We have investigated 
charged-particle production in ${d}$ + Au collisions at
$\sqrt{s_{NN}}$= 200 GeV as function of centrality
within the framework of an analytically soluble three-sources 
mode. Excellent agreement with 
recent PHOBOS pseudorapidity
distributions has been obtained. 
Only the midrapidity part (19\% in central collisions) of the
distribution function reaches equilibrium.
Although this fraction increases 
towards more peripheral collisions, the formation of a thermalized
parton plasma prior to hadronization can probably only be expected 
for central collisions.


\vfill\eject

\begin{thebibliography}{10}
\bibitem{bbb05}B.B. Back {\it et al.}, nucl-ex/0409021, {\it Phys. 
Rev. C}, 
in press.
\bibitem{wol99}G. Wolschin, {\it Eur. Phys. J.} $\bf{A5}$ (1999) 85.
\bibitem{wbs05}G. Wolschin, M. Biyajima,
T. Mizoguchi, and N. Suzuki, hep-ph/0503212;\\
submitted to {\it Phys. Lett. B}.
\bibitem{biy02}M. Biyajima, M. Ide, T. Mizoguchi, and N. Suzuki,\\
{\it Prog. Theor. Phys.} $\bf{108}$ (2002) 559; $\bf{109}$ (2003) 151.
\bibitem{wol03}G. Wolschin, {\it Phys. Lett.} $\bf{B569}$ (2003) 67;
{\it Phys. Rev.} $\bf{C69}$, 024906 (2004).
\bibitem{bha53}H.J. Bhabha, {\it Proc. Roy. Soc. (London)} $\bf{A 219}$ 
 (1953) 293.
\bibitem{nag84}S. Nagamiya and M. Gyulassy, {\it Adv. Nucl. Phys.} $\bf{13}$,
 (1984) 201.
\bibitem{ryb03}M. Rybczy\'nski, Z. W{\l}odarczyk, and G. Wilk,\\ {\it Nucl. 
Phys. (Proc. Suppl.)} $\bf{B122}$ (2003) 325.
\bibitem{wols99}G. Wolschin, {\it Europhys. Lett.} $\bf{47}$ (1999) 30.
\bibitem{biy04}M. Biyajima, M. Ide, M. Kaneyama, T. Mizoguchi, and 
N. Suzuki,\\
{\it Prog. Theor. Phys. Suppl.} $\bf{153}$ (2004) 344.
\bibitem{bbb04}B.B. Back {\it et al.}, {\it Phys. Rev. Lett.} $\bf{93}$ 
(2004) 082301.
\end{thebibliography}
\end{document}